\documentclass[11pt,a4paper]{article}

\usepackage{jheppub}
\usepackage{graphicx}
\usepackage{dcolumn}
\usepackage{bm}
\usepackage{amsmath}
\usepackage{braket}
\usepackage{slashed}    
\usepackage{epstopdf}
\usepackage{placeins}
\usepackage{multirow}
\usepackage{makecell}
\usepackage[shortlabels]{enumitem}
\usepackage{cleveref}
\usepackage{xcolor}
\usepackage[font=small, labelfont=bf, textfont={small,it}]{caption}

\newcommand{\beq}{\begin{eqnarray}}
\newcommand{\eeq}{\end{eqnarray}}
\newcommand{\beqnn}{\begin{eqnarray*}}
\newcommand{\eeqnn}{\end{eqnarray*}}
\newcommand{\tzero}{t_{\scriptscriptstyle{0}}}
\newcommand{\tone}{t_{\scriptscriptstyle{1}}}
\newcommand{\wzero}{w_{\scriptscriptstyle{0}}}
\newcommand{\TEK}{{\scriptscriptstyle{\rm TEK}}}
\newcommand{\NSVZ}{{\scriptscriptstyle{\rm NSVZ}}}
\newcommand{\W}{{\scriptscriptstyle{\rm W}}}
\newcommand{\Mgg}{M_{\scriptscriptstyle{\mathrm{g}\tilde{\mathrm{g}}}}}
\newcommand{\Ogg}{\mathcal{O}_{\scriptscriptstyle{\mathrm{g}\tilde{\mathrm{g}}}}}
\newcommand{\Cgg}{C_{\scriptscriptstyle{\mathrm{g}\tilde{\mathrm{g}}}}}
\newcommand{\SU}{\mathrm{SU}}
\newcommand{\Tr}{\mathrm{Tr}}
\newcommand{\Pf}{\mathrm{Pf}}
\newcommand{\NS}{\mathcal{N}}

\begin{document}

\title{The mass of the gluino-glue bound state in large-$N$ $\mathcal{N}=1$ Supersymmetric Yang--Mills theory}

\author[a]{Claudio Bonanno,}
\author[a]{Margarita Garc\'ia P\'erez,}
\author[a,b]{Antonio Gonz\'alez-Arroyo,}
\author[c,d]{\\Ken-Ichi Ishikawa,}
\author[d]{Masanori Okawa}

\affiliation[a]{Instituto de F\'isica T\'eorica UAM-CSIC, Calle Nicol\'as Cabrera 13-15,\\Universidad Aut\'onoma de Madrid, Cantoblanco, E-28049 Madrid, Spain}

\affiliation[b]{Departamento de F\'isica Te\'orica, Universidad Aut\'onoma de Madrid,\\M\'odulo 15, Cantoblanco, E-28049 Madrid, Spain}

\affiliation[c]{Core of Research for the Energetic Universe,\\Graduate School of Advanced Science and Engineering,\\Hiroshima University, Higashi-Hiroshima, Hiroshima 739-8526, Japan}

\affiliation[d]{Graduate School of Advanced Science and Engineering, Hiroshima University,\\Higashi-Hiroshima, Hiroshima 739-8526, Japan}

\emailAdd{claudio.bonanno@csic.es}
\emailAdd{margarita.garcia@csic.es}
\emailAdd{antonio.gonzalez-arroyo@uam.es}
\emailAdd{ishikawa@theo.phys.sci.hiroshima-u.ac.jp}
\emailAdd{okawa@hiroshima-u.ac.jp}

\abstract{We provide a first-principles non-perturbative determination of the mass of the lightest gluino-gluon bound state (gluino-glue) in large-$N$ $\mathcal{N}=1$ Supersymmetric Yang--Mills theory by means of numerical Monte Carlo simulations of the lattice-discretized theory, and exploiting large-$N$ twisted volume reduction. Our large-$N$ determination is consistent with naive extrapolation of previously-known $\SU(2)$ and $\SU(3)$ results.}

\keywords{$1/N$ Expansion, Lattice Quantum Field Theory, Other Lattice Field Theory, Supersymmetric Gauge Theory}

\maketitle

\section{Introduction}
In trying to understand many of the non-perturbative phenomena of gauge theories, particular relevance attain those theories which allow the simplification of some features. Supersymmetric (SUSY) gauge field theories~\cite{Wess:1974tw, Weinberg:2000cr} belong to this group as the cancellation of certain quantum effects among fermionic and bosonic degrees of freedom enable strong predictions and even exact results. On the other hand, other methodologies like the large-$N$ limit of gauge theories lead also to important simplifications such as factorization and the suppression of non-planar diagrams~\cite{tHooft:1973alw, Witten:1979kh}. Frequently supersymmetry and large $N$ appear as ingredients in simplifying the connection with string theory and other methodologies like holography~\cite{Maldacena:1997re, Aharony:1999ti}. 

In this paper we address the study of $\NS=1$ Supersymmetric Yang--Mills theory in the large-$N$ limit. Our methodology uses the formulation of this theory on the lattice with Wilson fermions~\cite{Curci:1986sm,Montvay:1995ea} together with the use of volume independence~\cite{PhysRevLett.48.1063,BHANOT198247,GONZALEZARROYO1983174,PhysRevD.27.2397,Kovtun:2007py,Unsal:2008ch} to attain the large-$N$ limit results in the most efficient way. Although the lattice formulation breaks Supersymmetry, it is well-known how one can approach the SUSY theory in the continuum limit by tuning the gluino mass to zero~\cite{Kaplan:1983sk,Curci:1986sm}.

To be more specific, our formulation makes use of twisted boundary conditions~\cite{tHooft:1979rtg} to achieve volume independence~\cite{GONZALEZARROYO1983174,PhysRevD.27.2397,Gonzalez-Arroyo:2010omx}. This has been tested extensively recently for pure Yang--Mills theory~\cite{Gonzalez-Arroyo:2012euf,Gonzalez-Arroyo:2014dua,GarciaPerez:2014azn,Gonzalez-Arroyo:2015bya,Perez:2017jyq,Perez:2020vbn,Bonanno:2023ypf}, but an extension to theories having fermions in the adjoint representation is possible and has been tested as well~\cite{Gonzalez-Arroyo:2013bta,GarciaPerez:2015rda,Butti:2022sgy,Bonanno:2024bqg}. The case of $\NS=1$ Supersymmetric Yang--Mills has also been studied recently. In particular in Ref.~\cite{Butti:2022sgy} we studied the scale setting of the theory, a necessary ingredient to take the continuum limit of the lattice theory. This was done for the theory with non-zero gluino mass, but also extrapolated to the supersymmetric zero gluino mass case. The results showed a $\beta$-function in agreement with the prediction of the continuum supersymmetric theory. In a later work we were able to evaluate the gluino condensate of the theory from our first principles approach~\cite{Bonanno:2024bqg}, elucidating a long-time puzzle about its $N$-dependence in favor of the weak coupling prediction~\cite{Novikov:1985ic,Davies:1999uw,Anber:2024mco}.

As mentioned earlier, our methodology is based on twisted reduced models. For the pure gauge part this is given by the twisted Eguchi--Kawai model~\cite{PhysRevD.27.2397}. Adjoint fermions are not subleading in the large-$N$ limit as for the case of a finite number of fundamental flavours, and demand standard dynamical fermion techniques to generate gauge ensembles. Full details can be obtained in the following section and from the two previously mentioned publications. The twisted volume reduction technique, although very efficient to compute single trace observables in the large-$N$ and large-volume limit, has limitations concerning other observables. In particular, the space-time dependence of the correlation of two single-trace observables is lost. This makes it impossible to compute the glueball spectrum and also masses of flavour singlet mesons. 
However, this is not the case for the gluino-glue bound state that is studied in this paper. It should be noted, however, that because of Supersymmetry this state should be degenerate in mass with the remaining states in the supermultiplet. Thus, we can obtain the mass of this supermultiplet and compare it with the computations done at finite $N$.

These computations are scarce because of the difficulty inherent in such a dynamical fermion generation. We should mention here the work of the DESY--M\"unster--Jena--Regensburg collaboration, which have pursued a systematic program targeting the study of several properties of supersymmetric $\SU(2)$ and $\SU(3)$ Yang--Mills theories from the lattice~\cite{Munster:2014bea,Munster:2014cja,Bergner:2015adz,Ali:2018fbq,Ali:2018dnd,Ali:2019agk,Ali:2019gzj,Bergner:2019dim,Bergner:2024ttq}, see also the following recent reviews~\cite{Bergner:2016sbv,Bergner:2022snd,Schaich:2022xgy}. Recently, they were able to show for the $\SU(2)$~\cite{Ali:2019gzj} and $\SU(3)$~\cite{Ali:2019agk} theories the formation of the lightest supermultiplet in the SUSY-restoring chiral-continuum limit. In this limit the lightest $0^{++}$ state (mixing of the lightest glueball and of the adjoint $f_0$ meson), the lightest flavor-singlet meson (adjoint $\eta^\prime)$ and the lightest gluino-glue bound state become degenerate in mass, thus pointing out that Supersymmetry is unbroken (Supersymmetry restoration was also established by the same authors from the lattice study of supersymmetric Ward identities~\cite{Ali:2020mvj}). Our work can be placed in this context, and constitutes a first step towards the study of the realization of Supersymmetry in the large-$N$ SUSY gauge theory.

This paper is organized as follows. In Sec.~\ref{sec:numerical_setup} we summarize our twisted volume-reduced lattice setup, and the numerical techniques adopted to extract the gluino-glue mass. In Sec.~\ref{sec:results} we present our large-$N$ results for the gluino-glue mass, presenting our extrapolations towards the thermodynamic and chiral-continuum SUSY limit, and comparing it with previous finite-$N$ results. Finally, in Sec.~\ref{sec:conclu} we draw our conclusions and discuss future outlooks of this study.

\section{Numerical setup}\label{sec:numerical_setup}

In this section we briefly review our numerical lattice setup, which has also been employed in our previous numerical investigations of large-$N$ SUSY Yang--Mills theory~\cite{Butti:2022sgy,Bonanno:2024bqg}, and we discuss how to compute the gluino-glue mass in the reduced model.

\subsection{Lattice action}\label{sec:lat_setup}

We inherited gauge configurations from the previous study~\cite{Butti:2022sgy}, whose setup is summarized in the following. The Twisted Eguchi--Kawai (TEK) partition function of Super Yang--Mills in the path-integral formulation reads:
\beq
Z_{\TEK} = \int [\mathrm{d}U] \Pf\left\{C D_{\W}^{(\TEK)}[U]\right\}\mathrm{e}^{-S_{\TEK}[U]} \,,
\eeq
where $S_{\TEK}$ is the gauge action, $D_{\W}^{(\TEK)}$ is the lattice Wilson--Dirac operator in the adjoint representation, and the matrix $C$ is the charge-conjugation operator, defined so as to satisfy the following relations: $\gamma_\mu^{\rm t} C = - \gamma_\mu C$ and $C^{\rm t}=-C$. $\Pf$ stands for the Pfaffian of the anti-symmetric matrix $CD_{\W}^{(\TEK)}$.

Our large-$N$ discretization is rooted on the idea of large-$N$ volume independence~\cite{PhysRevLett.48.1063}. More precisely, by virtue of large-$N$ twisted volume reduction~\cite{GONZALEZARROYO1983174,PhysRevD.27.2397}, the lattice large-$N$ $\SU(N)$ Yang--Mills theory is equivalent to a matrix model. The TEK Wilson plaquette gauge action, defined in terms of the $d=4$ gauge links $U_\mu \in \SU(N)$ (transforming under the fundamental representation of the gauge group), reads:
\beq
S_{\TEK}[U] = b N \sum_{\mu \, = \, 1}^{d}\sum_{\nu \,\ne \, \mu}\Tr\left\{\mathrm{I} - z_{\mu\nu}^{*}U_\mu U_\nu U_\mu^\dagger U_\nu^\dagger\right\},
\eeq
where $\mathrm{I}$ is the $N \times N$ identity matrix, $b=1/\lambda_{\scriptscriptstyle{\rm L}}$ is the inverse bare 't Hooft coupling, and $z_{\mu\nu}$ is the twist factor. This action can thus be interpreted as the standard Wilson lattice gauge action defined on a single space-time point with twisted boundary conditions.

Several possible choices for the twist are possible. In this work we chose the \emph{symmetric twist}, that takes the form:
\beq\label{eq:twist_factor}
z_{\mu\nu}=\exp\left\{\mathrm{i} \, 2\pi n_{\mu\nu}/N \right\}, \quad n_{\mu\nu}= L \, k(N), \qquad \qquad (k,L\in\mathbb{N}, \,\,\,\, \nu>\mu),
\eeq
where $N=L^2$ is taken to be a perfect square, and where $n_{\mu\nu}$ satisfies $n_{\mu\nu}=-n_{\nu\mu}$. As explained in Ref.~\cite{Gonzalez-Arroyo:2010omx}, the integer number $k$ must be taken co-prime with $L=\sqrt{N}$ and must be increased as $N$ increases in order to avoid center-symmetry breaking~\cite{Ishikawa:2003,Bietenholz:2006cz,Teper:2006sp,Azeyanagi:2007su}. Moreover, non-planar finite-$N$ corrections are greatly reduced with appropriate choices of $k$, see Refs.~\cite{Gonzalez-Arroyo:2010omx,Gonzalez-Arroyo:2014dua,Chamizo:2016msz,GarciaPerez:2018fkj,Perez:2017jyq,Bribian:2019ybc}.

The Wilson--Dirac operator for adjoint fermions in the TEK formulation was derived in Ref.~\cite{Gonzalez-Arroyo:2013bta} and reads:
\beq
D_{\W}^{(\TEK)}[U] = \frac{1}{2 \kappa}\mathrm{I} - \frac{1}{2}\sum_{\mu \, = \, 1}^{d}\bigg[(\mathrm{I}-\gamma_\mu)U_\mu^{{\scriptscriptstyle{(\rm adj)}}} + (\mathrm{I}+\gamma_\mu)\left(U_\mu^{{\scriptscriptstyle{(\rm adj)}}}\right)^{\rm t} \bigg],
\eeq
with $\kappa$ the Wilson hopping parameter, and $U_\mu^{{\scriptscriptstyle{(\rm adj)}}}$ the $\SU(N)$ gauge links in the adjoint representation.

Unlike in the continuum theory, the Pfaffian of the lattice Wilson--Dirac operator is not guaranteed to be positive. To treat this (mild) sign problem, we follow the strategy extensively used in lattice simulations of SUSY Yang--Mills theories by the DESY--Jena--Regensburg--M\"unster collaboration~\cite{Munster:2014bea,Munster:2014cja,Bergner:2015adz,Ali:2018fbq,Ali:2018dnd,Ali:2019agk,Ali:2019gzj,Bergner:2019dim,Bergner:2024ttq}, namely, to perform sign-quenched (sq) simulations according to the following probability distribution,
\beq
\mathcal{Z}_{\TEK}^{(\rm sq)} = \int [\mathrm{d}U] \left\vert \Pf\left\{C D_{\W}^{(\TEK)}[U]\right\} \right\vert \mathrm{e}^{-S_{\TEK}[U]},
\eeq
incorporating the sign of the Pfaffian into the observable via a standard reweighting procedure:
\beq
\braket{\mathcal{O}} = \frac{\left\langle \mathcal{O} \, \mathrm{sign}\left[\Pf\left\{C D_{\W}^{(\TEK)}[U]\right\}\right]\right\rangle_{\rm sq}}{\left\langle\mathrm{sign}\left[\Pf\left\{C D_{\W}^{(\TEK)}[U]\right\}\right]\right\rangle_{\rm sq}}.
\eeq
Sign-quenched simulations are free of any sign problem, thus sign-quenched expectation values can be computed from the sampling of the sign-quenched partition function using standard Monte Carlo algorithms. In our work we used the Rational Hybrid Monte Carlo (RHMC) algorithm~\cite{Kennedy:1998cu,Clark:2003na,Clark:2004cp,Clark:2005sq,Clark:2006fx,Clark:2006wp}, whose details for the simulation of the TEK model in the presence of adjoint fermions are described in~\cite{Butti:2022sgy}.

The occurrence of negative signs of the Pfaffian can be detected by counting the negative real eigenvalues of the Wilson--Dirac operator. In our previous study~\cite{Butti:2022sgy} we did not find any occurrence of such negative eigenmodes, meaning that the Pfaffian was positive for every generated trajectory, and thus no reweighting at all was needed at the end.

Finally, it is important to recall the nature of finite-$N$ corrections in the TEK model. Twisted volume reduction allows to gain information about the $N=\infty$ value of physical observables. In practice, we are forced to perform simulations for $N$ very large but finite. Finite-$N$ corrections in the TEK model---which have been studied in detail both perturbatively and non-perturbatively in Refs.~\cite{Gonzalez-Arroyo:2014dua,Perez:2017jyq}---will mainly manifest themselves as finite-volume corrections, as if the results were obtained for a finite box with effective size $\ell = a L = a \sqrt{N}$, with $a$ the lattice spacing. The \emph{physical} $N$-dependence of any quantity cannot be obtained from the TEK model alone, but only combining $N=\infty$ TEK results with those obtained on standard extended lattices for finite values of $N$. In Sec.~\ref{sec:finiteN_effects} we will show and compare results obtained from the TEK model for 3 values of $N$, $N=169$, 289 and $N=361$ [obtained choosing respectively $k(N)=5$, 5 and $7$ for the twist factor in Eq.~\eqref{eq:twist_factor}], in order to assess the magnitude of finite-volume effects affecting the determination of the gluino-glue mass.

\subsection{Extraction of the gluino-glue mass in the reduced model}

Following the ideas put forward in previous studies of SUSY spectra in Refs.~\cite{Donini:1997hh,Ali:2018dnd,Ali:2019agk,Ali:2019gzj}, we obtained the gluino-glue masses from the exponential decay of the Euclidean-time correlator
\beq\label{eq:full_tcorr}
\Cgg^{\alpha\beta}(\tau) = \int \mathrm{d}^3 x \braket{\Ogg^{\alpha}(\tau,\vec{x})\Ogg^{\beta}(0,\vec{0})},
\eeq
with the following interpolating spatial operator:
\beq
\Ogg^{\alpha} = \sigma^{\alpha\beta}_{ij} \, \Tr\left(G_{ij}\lambda^{\beta} \right), \qquad \qquad \sigma_{ij} = \frac{\mathrm{i}}{2} [\gamma_i,\gamma_j],
\eeq
where $G_{ij}$ is the gluon field strength in the spatial directions, $\lambda$ is the gluino field, and where the trace is performed over color indices. The correlator in Eq.~\eqref{eq:full_tcorr} admits the following decomposition~\cite{Donini:1997hh}:
\beq\label{eq:corr_gluinoglue}
\Cgg^{\alpha\beta} = \frac{1}{4} C_1 \delta^{\alpha\beta} + \frac{1}{4}C_2 \gamma_4^{\alpha\beta},
\eeq
\beq
C_1 = \Cgg^{\alpha\beta} \delta^{\beta\alpha}, \qquad C_2 = \Cgg^{\alpha\beta} \gamma_4^{\beta\alpha},
\eeq
where both the even $C_1$ and the odd $C_2$ correlators lead to the same mass. After verifying this expectation in a few cases, we always used the correlator $C_1$, which always turned out to be less noisy.

In the reduced model, temporal correlators are obtained anti-transforming from Fourier space the zero-spatial-momentum correlator~\cite{Gonzalez-Arroyo:2015bya,Perez:2020vbn,Butti:2022sgy}:
\beq
C_1(\tau) = \sum_{\omega} \mathrm{e}^{- \mathrm{i} \, \omega \tau} \, \widetilde{C}_1(\omega).
\eeq
We chose the temporal extent to be $T = 2 a \sqrt{N}$ so that the temporal component $\omega$ of the momentum takes values:
\beq
a \omega_m = \frac{\pi}{\sqrt{N}} m, \qquad \qquad m\in\mathbb{N}.
\eeq
The temporal correlator in momentum space is given by:
\beq\label{eq:latcorr}
\widetilde{C}_1(\omega) = \frac{1}{4N^2} \sigma_{ij}^{\alpha\beta}\left\langle\Tr\left[P_{ij} S^{\beta\gamma}(\omega)P_{lm} \right]\right\rangle \sigma_{lm}^{\gamma\alpha}.
\eeq
Here $P_{\mu\nu}$ is the clover-discretization of the gluon field strength $G_{\mu\nu}$,
\beq
P_{\mu\nu} \equiv \frac{1}{8 \mathrm{i}}\left(\mathrm{Clov}_{\mu\nu} - \mathrm{Clov}_{\mu\nu}^{\dagger}\right) - \frac{1}{8N\mathrm{i}}\Tr\left(\mathrm{Clov}_{\mu\nu} - \mathrm{Clov}_{\mu\nu}^{\dagger}\right),
\eeq
\beq
\mathrm{Clov}_{\mu\nu} \equiv z^*_{\mu\nu}\left(U_\mu U_\nu U_\mu^\dagger U_\nu^\dagger + U_\mu^\dagger U_\nu^\dagger U_\mu U_\nu + U_\nu^\dagger U_\mu U_\nu U_\mu^\dagger + U_\nu U_\mu^\dagger U_\nu^\dagger U_\mu\right),
\eeq
while $S(\omega)$ is the lattice gluino propagator in Fourier space, obtained from the inversion of the Fourier-space lattice Wilson--Dirac operator:
\beq
\begin{aligned}
S^{-1}(\omega) &= \widetilde{D}_{\W}^{(\TEK)}(\omega) \\
\\[-1em]
&= \frac{1}{2 \kappa}\mathrm{I} - \frac{1}{2}\sum_{\mu \, = \, 1}^{d}\bigg[(\mathrm{I}-\gamma_\mu)U_\mu^{{\scriptscriptstyle{(\rm adj)}}} \, e^{\mathrm{i} \, p_\mu} + (\mathrm{I}+\gamma_\mu)\left(U_\mu^{{\scriptscriptstyle{(\rm adj)}}} \, e^{\mathrm{i} \, p_\mu} \right)^\dagger \bigg], \,\,\, p_\mu=(\omega,\vec{0}).
\end{aligned}
\eeq
More technical details on the inversion of the Dirac operator can be found in the Appendix of Ref.~\cite{Butti:2022sgy}.

In order to improve the overlap of the interpolating operator with the desired state, on the lattice it is customary to identify the best-overlapping operator from the resolution of a Generalized Eigen-Value Problem (GEVP)~\cite{Berg:1982kp,Michael:1985ne,Luscher:1990ck}:
\beq \label{eq:gevp}
C_1^{ab}(\tau)v_{b} = \lambda(\tau,\tau_0) C_1^{ab}(\tau_0) v_{b},
\eeq
\beq
C_1^{ab}(\tau) = \sum_{\omega} \mathrm{e}^{- \mathrm{i} \, \omega \tau} \, \widetilde{C}^{ab}_1(\omega), \qquad \widetilde{C}^{ab}_1(\omega) = \frac{1}{4N^2} \sigma_{ij}^{\alpha\beta}\left\langle\Tr\left[P^a_{ij} S^{\beta\gamma}(\omega)P^b_{lm} \right]\right\rangle \sigma_{lm}^{\gamma\alpha}.
\eeq
In the case at hand the indices $(a,b)$ run over different levels of spatial stout smearing~\cite{PhysRevD.69.054501} of the gluon fields entering the clover-discretized gluon strengths $P_{\mu\nu}$. Its practical implementation is similar to that of spatial APE smearing~\cite{APE:1987ehd} (introduced in~\cite{Gonzalez-Arroyo:2012euf} and described in~\cite{Butti:2022sgy} for reduced models), the main difference being that the $\SU(N)$ projection is achieved via an exponentiation of the spatial staple, rather than via a linear transformation. The correlator of the best-overlapping operator is obtained from the eigenvector $\overline{v}$ corresponding to the largest eigenvalue $\overline{\lambda}(\tau,\tau_0)$:
\beq
C_1^{({\scriptscriptstyle{\rm best}})}(\tau) \equiv C_1^{ab}(\tau)\overline{v}_a^{*} \overline{v}_b \underset{\tau\to\infty}{\sim} \exp\left(-\Mgg \tau\right).
\eeq
In all cases we solved the GEVP for fixed values of $\tau_0/a=1$ and $\tau=\tau_0 + a$, and we also verified in a few cases that $\tau_0/a=2$ gave compatible results. Once the best-overlapping operator is obtained, the gluino-glue mass $\Mgg$ is extracted from an exponential best fit of $C_1^{({\scriptscriptstyle{\rm best}})}(\tau)$ to:
\beq\label{eq:expfit_mass}
C_1^{({\scriptscriptstyle{\rm best}})}(\tau) = A \cosh\left[\Mgg (\tau -T/2) \right], \qquad \tau \in [\tau_1,\tau_2].
\eeq
To validate the precision of our mass determinations, we conducted an independent calculation by solving the GEVP for all values of $\tau$ larger than $\tau_0$.  In the limit where the correlator is predominantly influenced by a single exponential, the largest eigenvalue of Eq.~\eqref{eq:gevp}  is expected to exhibit the following behaviour:
\beq
\overline{\lambda}(\tau,\tau_0) \underset{\tau,\tau_0\to\infty}{\sim} \exp\left[-\Mgg(\tau-\tau_0)\right].
\eeq
In alignment with this expectation, an effective mass was derived from the plateaux observed in:
\beq\label{eq:eff_mass}
\Mgg^{({\scriptscriptstyle{\rm eff}})}(\tau) \equiv - \frac{ \log\left[\overline{\lambda}(\tau,\tau_0)\right]}{\tau-\tau_0},
\eeq
which occurred within the same range $\tau \in[\tau_1,\tau_2]$ where $C_1^{({\scriptscriptstyle{\rm best}})}(\tau)$ could be accurately fitted via Eq.~\eqref{eq:expfit_mass}. Both determinations of the gluino-glue mass yielded compatible results.

\section{Results}\label{sec:results}

In this section we present our results for the gluino-glue mass as a function of $N$, of the lattice spacing and of the gluino mass. We will start, in the next subsection, describing the extrapolation towards the thermodynamic limit ($N=\infty$). In the following one, we will study the gluino-glue mass in the SUSY-restorting limit (chiral-continuum), necessary to obtain our final $\SU(\infty)$ determination given there. Finally, in the last subsection we discuss the consistency of our result with previous ones for finite values of $N$.

\subsection{Gluino-glue mass determination and \texorpdfstring{$N=\infty$}{N = infinity} thermodynamic limit}\label{sec:finiteN_effects}

\begin{figure}[!t]
\centering
\includegraphics[scale=0.9]{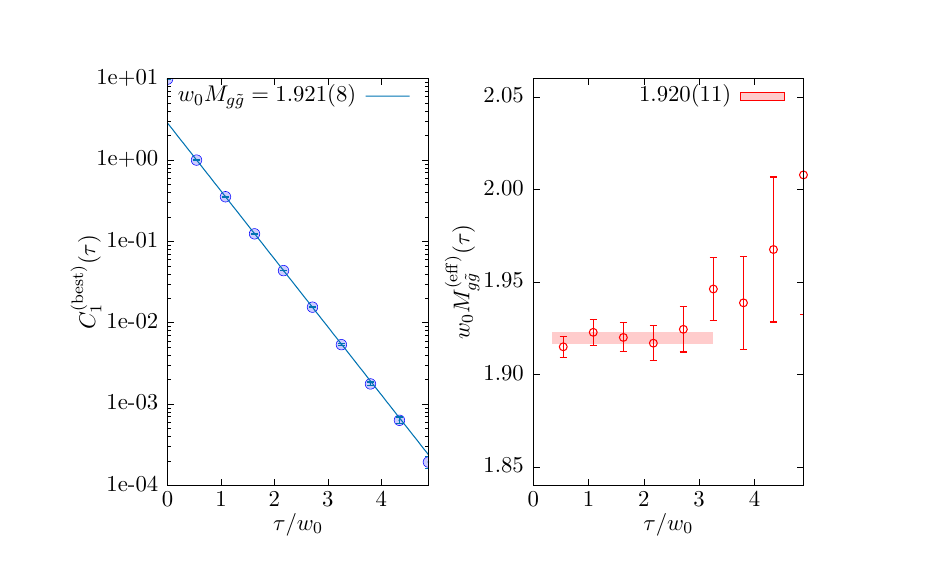}
\caption{Left panel: example of the exponential fit to the gluino-glue best overlapping correlator in the range $0.5<\tau/\wzero<3$ according to~\eqref{eq:expfit_mass}. Right panel: constant fit to the plateau in the effective mass~\eqref{eq:eff_mass} observed in the same range of Euclidean times. Both fits give perfectly agreeing results for the gluino-glue mass. Plots refer to $N=361$, $b=0.345$, $\kappa=0.1800$.}
\label{fig:corr_ex}
\end{figure}

All our determinations of the gluino-glue mass $\Mgg$ are reported in Tab.~\ref{tab:summary_raw_results}. We report results for several values of the bare parameters $(b,\kappa)$ and of the number of colors $N$. An example of the fit to the correlator used to obtain the masses is shown in Fig.~\ref{fig:corr_ex}.

\begin{table}[!t]
\small
\begin{center}
\begin{tabular}{|c|c|c|c||c|c|c|}
\hline
$b$ & $\kappa$ & $a m_{\pi}$ & $a/\sqrt{8\tone}$ & \makecell{$a\Mgg$\\$N=169$} & \makecell{$a\Mgg$\\$N=289$} & \makecell{$a\Mgg$\\$N=361$} \\
\hline
\multirow{4}{*}{0.340} & 0.1850 & 0.977(4) & 0.3469(70) & 1.073(13) & 1.0328(49) & 1.0326(32) \\
& 0.1875 & 0.819(7) & 0.3137(89) & 0.944(13) & 0.8929(43) & 0.8899(58) \\
& 0.1890 & 0.719(4) & 0.2867(72) & 0.861(12) & 0.806(15) & 0.7934(56) \\
& 0.1910 & 0.540(5) & 0.2474(36) & 0.814(14) & 0.6860(79) & 0.6494(59) \\
\hline
\multirow{4}{*}{0.345} & 0.1800 & 1.043(6) & 0.3180(94) &   & 1.0533(50) & 1.0427(44) \\
& 0.1840 & 0.821(5) & 0.2743(74) &   & 0.8742(73) & 0.8570(48) \\
& 0.1868 & 0.631(6) & 0.2340(30) &   & 0.7374(90) & 0.7138(72) \\
& 0.1896 & 0.353(5) & 0.1781(64) &   & 0.609(27) & 0.577(16) \\
\hline
\multirow{4}{*}{0.350} & 0.1775 & 1.001(7) & 0.2676(74) & 1.0553(92) & 0.9887(47) & 0.9880(47) \\
& 0.1800 & 0.883(7) & 0.2498(68) & 0.9535(88) & 0.8946(54) & 0.8852(44) \\
& 0.1825 & 0.733(6) & 0.2214(31) & 0.812(27) & 0.791(13) & 0.7609(80) \\
& 0.1850 & 0.540(6) & 0.1879(23) &  & 0.653(15) & 0.651(14) \\
\hline
\end{tabular}
\end{center}
\caption{Summary of the finite-$N$ determinations of the gluino-glue mass as a function of the bare parameters $b$ and $\kappa$. Results for the adjoint-pion mass $am_\pi$ and for the lattice spacing $a/\sqrt{8\tone}$ come from Ref.~\cite{Butti:2022sgy}.}
\label{tab:summary_raw_results}
\end{table}

In addition, Tab.~\ref{tab:summary_raw_results} also contains the values of the lattice spacing $a$, and of the adjoint-pion mass $m_\pi$, as determined in our previous study~\cite{Butti:2022sgy}. These data will be instrumental to extrapolate our finite-lattice-spacing and finite-gluino-mass results for the gluino-glue mass towards the SUSY-restoration limit, i.e., chiral-continuum limit~\cite{Kaplan:1983sk,Curci:1986sm}, reached for $a=0$ and $m_\pi=0$~\cite{Munster:2014bea,Munster:2014cja}. We emphasize that, as explained in~\cite{Butti:2022sgy}, this unphysical pion is not part of the actual SUSY spectrum, and can only be defined adding a second quenched gluino to the theory. However, according to the framework of partially-quenched chiral perturbation theory, its mass can be used to tune the bare gluino mass towards the chiral limit~\cite{Munster:2014bea,Munster:2014cja}.

In order to set the absolute scale, we expressed the lattice spacing $a$ in terms of the reference hadronic gradient flow scale $\sqrt{8t_1}$. Gradient flow~\cite{Narayanan:2006rf,Luscher:2009eq,Lohmayer:2011si} is a smoothing procedure that evolves the gauge fields according to the flow-time equations:
\beq
\partial_t B_\mu (t) = D_\nu G_{\mu \nu} (t), \qquad B_\mu (t = 0) = A_\mu.
\eeq
From the flow-time evolution of the action density
\beq
E(t) = \frac{1}{2} \Tr \left \{G_{\mu \nu} (t)G_{\mu \nu} (t)\right\},
\eeq
one can define the reference scale $\sqrt{8t_1}$ as:
\beq
\frac{\braket{t^2 E(t)}}{N} \Bigg\vert_{t\,=\,{\tone}} &=& c \qquad \text{with} \qquad c = 0.05.
\eeq

We did not choose the popular scale $\sqrt{8\tzero}$, corresponding to $c=0.1$, as in our previous study~\cite{Butti:2022sgy} we verified $\tone$ to be less sensitive to finite-volume (i.e., finite-$N$) effects, being the cut $c$ of the energy density performed at a smaller flow time. Note also the factor of $N$, coming from the $N$-scaling of the energy density:
\beq\label{eq:ref_scale_t1}
\braket{t^2 E(t)} \underset{N\to\infty}{\sim} \frac{N^2-1}{N} \underset{N\to\infty}{\sim} N.
\eeq

In order to compare our large-$N$ results for the gluino-glue mass with previous results obtained for smaller values of $N$, it is also useful to introduce the following scale:
\beq\label{eq:ref_scale_w0}
t\,\frac{\mathrm{d}\textcolor{white}{t}}{\mathrm{d} t} \left[ \frac{t^2\braket{E(t)}}{N}\right]\Bigg\vert_{t\,=\,\wzero^2} = c, \qquad \text{with} \qquad c = 0.1.
\eeq
In Ref.~\cite{Butti:2022sgy} we determined the ratio:
\beq\label{eq:ratio_w0_t1_largeN}
\wzero/\sqrt{8\tone} = 0.586(10),
\eeq
thus converting our determinations of $\sqrt{8 \tone} \Mgg$ in units of $\wzero$ poses no difficulty. For completeness, we also quote~\cite{Butti:2022sgy}:
\beq\label{eq:ratio_t0_t1_largeN}
\sqrt{\tzero/\tone} = 1.627(50).
\eeq

As outlined in Sec.~\ref{sec:lat_setup}, in the TEK model finite-$N$ effects manifest themselves as finite-volume effects. In particular, finite-$N$ results must be understood as obtained on a finite box with effective size $\ell = a \sqrt{N}$, Thus, the $N$-dependence in the TEK model is rather an effective-volume dependence, and has nothing to do with the physical $N$-dependence of the observables, which can only be grasped by combining TEK large-$N$ and standard finite-$N$ results (see Sec.~\ref{sec:Ndep}). Thus, comparing results obtained for different values of $N$ allows to assess the magnitude of finite-volume effects affecting our calculation.

\begin{figure}[!t]
\centering
\includegraphics[scale=0.62]{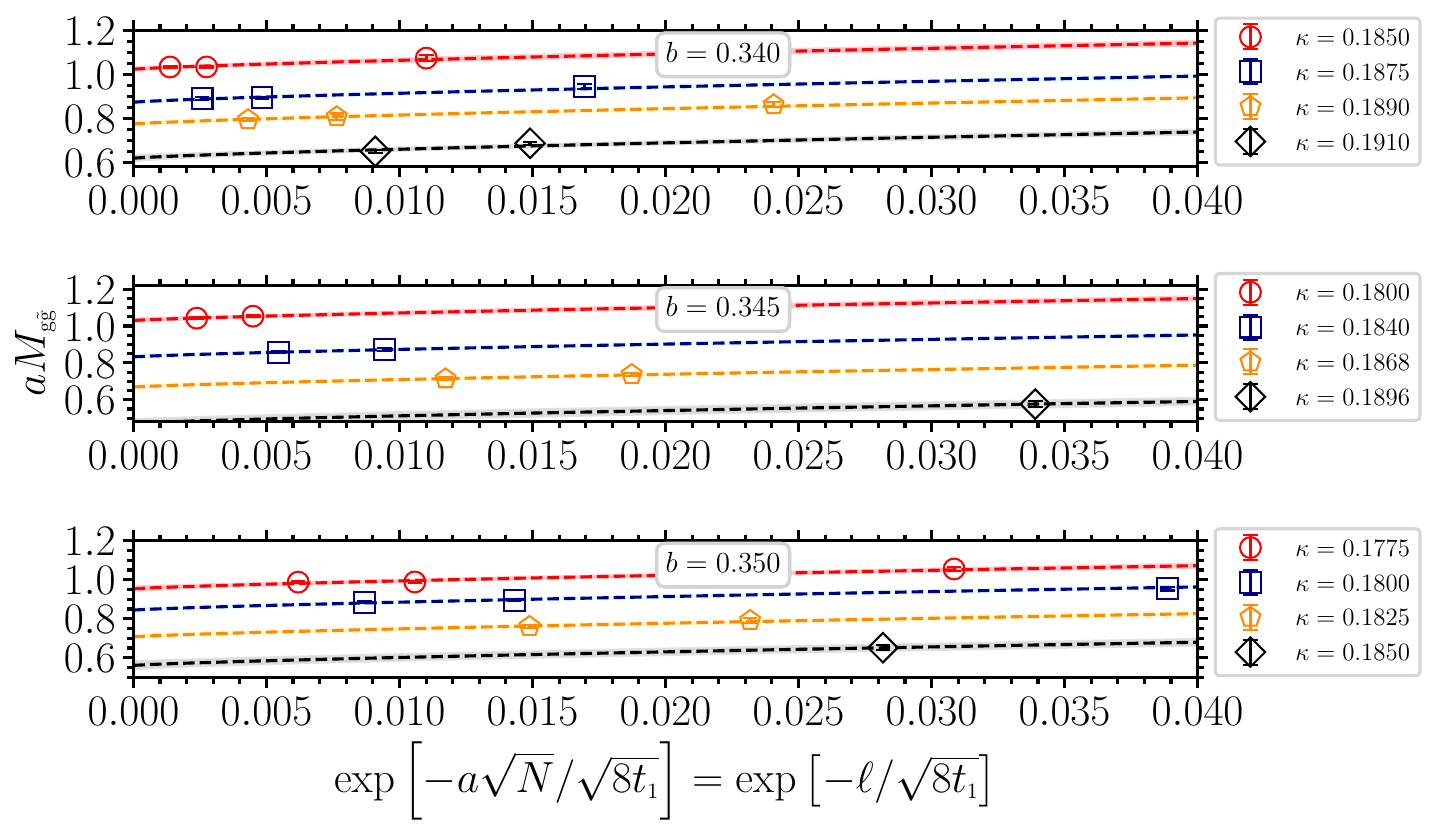}
\caption{Extrapolation towards the thermodynamic $N=\infty$ limit of our finite-volume determinations of $a\Mgg$ reported in Tab.~\ref{tab:summary_raw_results} according to fit function~\eqref{eq:fit_func}. The shown global best fit corresponds to $\exp[-\ell/\sqrt{8\tone}] < 0.04$, i.e., $\ell/\sqrt{8\tone}\gtrsim 3.2$. Best fit yields a reduced chi-squared of 1.08 with 13 degrees of freedom, corresponding to a $p$-value $\simeq 37\%$}
\label{fig:Ninf_extrap}
\end{figure}

Since a mild $N$-dependence can be observed in our determinations, cf.~Tab.~\ref{tab:summary_raw_results}, we extrapolated our results towards the thermodynamic $N=\infty$ limit according to the following fit function,
\beq\label{eq:fit_func}
(a\Mgg)(N,b,\kappa) = (a\Mgg)(\infty,b,\kappa) + A_1\exp\left[-A_2 \, \frac{\ell(N,b,\kappa)}{\sqrt{8\tone}}\right],
\eeq
\beq
\frac{\ell(N,b,\kappa)}{\sqrt{8\tone}} = \frac{a(b,\kappa)}{\sqrt{8\tone}} \times \sqrt{N},
\eeq
performing a global best fit to our data assuming that $A_1$ and $A_2$ are independent of the bare parameters $b$ and $\kappa$. This fit function is justified on the basis that finite-volume effects in the TEK model coming from planar diagrams are expected to be the same affecting a standard calculations performed on an extended lattice of size equal to the effective size $\ell = a \sqrt{N}$, which are known to be exponentially suppressed in the physical volume, at least when $\ell \gtrsim 1/\Lambda$ (with $\Lambda$ the dynamically-generated scale of the theory). On the other hand, power-law finite-$N$ corrections due to non-planar diagrams are expected to drop when reduction holds (in Ref.~\cite{Perez:2017jyq} this is analytically worked out in perturbation theory for Wilson loops).

We tried several best fits choosing different fit ranges, namely $\ell/\sqrt{8\tone}\gtrsim 2.8$, 3.0 and 3.2, and found that in all cases fit function~\eqref{eq:fit_func} perfectly described our data. As an example, in Fig.~\ref{fig:Ninf_extrap} we show the obtained best fit for $\ell/\sqrt{8\tone}\gtrsim 3.2$, yielding a reduced chi-squared of 1.08 with 13 degrees of freedom, corresponding to a $p$-value $\simeq 37\%$. The $N=\infty$ extrapolations of $a\Mgg$ and the free fit parameters $A_1,A_2$ changed negligibly upon varying the cuts on the values of the physical volume. In the end, in Tab.~\ref{tab:summary_Ninf_results} we quote as our final $N=\infty$ values of the gluino-glue mass those obtained excluding the smaller volumes, i.e., only fitting data satisfying $\ell\sqrt{8\tone}\gtrsim 3.2$, which yielded the most conservative statistical errors.

\begin{table}[!t]
\small
\begin{center}
\begin{tabular}{|c|c|c|c|c|}
\hline
$b$ & $\kappa$ & $a m_{\pi}$ & $a/\sqrt{8\tone}$ & \makecell{$a\Mgg$\\$N=\infty$} \\
\hline
\multirow{4}{*}{0.340} & 0.1850 & 0.977(4) & 0.3469(70) & 1.0221(69) \\
 & 0.1875 & 0.819(7) & 0.3137(89) & 0.8723(98) \\
 & 0.1890 & 0.719(4) & 0.2867(72) & 0.773(11) \\
 & 0.1910 & 0.540(5) & 0.2474(36) & 0.618(13) \\
 \hline
\multirow{4}{*}{0.345} & 0.1800 & 1.043(6) & 0.3180(94) & 1.0299(88) \\
 & 0.1840 & 0.821(5) & 0.2743(74) & 0.832(11) \\
 & 0.1868 & 0.631(6) & 0.2340(30) & 0.669(14) \\
 & 0.1896 & 0.353(5) & 0.1781(64) & 0.472(20) \\
 \hline
\multirow{4}{*}{0.350} & 0.1775 & 1.001(7) & 0.2676(74) & 0.953(12) \\
 & 0.1800 & 0.883(7) & 0.2498(68) & 0.844(12) \\
 & 0.1825 & 0.733(6) & 0.2214(31) & 0.707(15) \\
 & 0.1850 & 0.540(6) & 0.1879(23) & 0.561(19) \\
\hline
\end{tabular}
\end{center}
\caption{Summary of our final $N=\infty$ determinations of the gluino-glue mass in lattice units, corresponding to the thermodynamic extrapolations shown in Fig.~\ref{fig:Ninf_extrap}.}
\label{tab:summary_Ninf_results}
\end{table}

\subsection{The \texorpdfstring{$N=\infty$}{N = infinity} gluino-glue mass in the SUSY limit}

After extrapolating our results towards the $N=\infty$ thermodynamic limit, we now move towards the calculation of the gluino-glue mass in the SUSY (chiral-continuum) limit. To this end, we first computed the following scaling quantity using the results for $a\Mgg(N=\infty)$ reported in Tab.~\ref{tab:summary_Ninf_results}:
\beq
\frac{\wzero}{\sqrt{8\tone}} \times \frac{\sqrt{8\tone}}{a(b,\kappa)} \times (a\Mgg)(\infty,b,\kappa) = (\wzero \Mgg)(\infty,b,\kappa).
\eeq
We chose to express the gluino-glue mass in terms of the reference scale $\wzero$ in order to allow comparison with previous lattice results obtained in the literature for $N=2,3$, which are expressed in terms of this scale.

\begin{figure}[!t]
\centering
\includegraphics[scale=0.62]{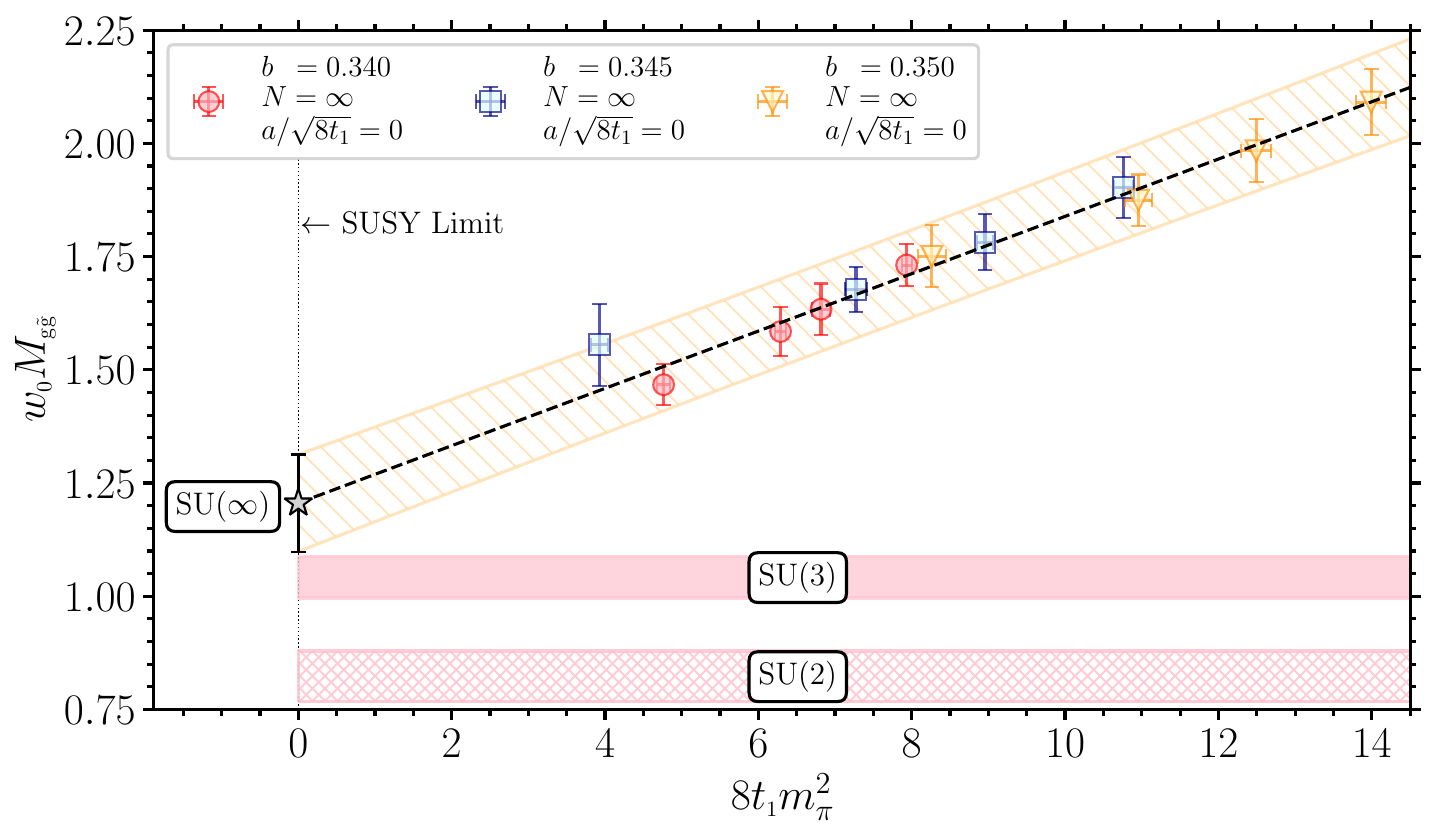}
\caption{Extrapolation towards the SUSY limit of the $N=\infty$ gluino-glue mass in units of the reference hadronic scale $\wzero$ according to the fit function~\eqref{eq:SUSY_limit_fit_function}. We obtain a reduced chi-squared of about $0.32$ with 9 degrees of freedom. Plot shows the continuum best-fit curve, i.e., the curve obtained from the best fit of the data to Eq.~\eqref{eq:SUSY_limit_fit_function} with $c_1 = 0$. Also the depicted points correspond to continuum extrapolated determinations, and are found subtracting the quantity $c_1a/\sqrt{8\tone}$ to the finite-lattice-spacing results of Tab.~\ref{tab:summary_Ninf_results}. Our $\SU(\infty)$ result in the SUSY limit (corresponding to 0 in the horizontal scale) is compared to the $\SU(2)$ and $\SU(3)$ determinations of Refs.~\cite{Bergner:2014ska,Ali:2019gzj,Ali:2019agk}, also obtained in the SUSY limit, see also the text in Sec.~\ref{sec:Ndep}.}
\label{fig:SUSY_limit}
\end{figure}

Then, we performed a simultaneous continuum and chiral extrapolation according to the following fit function, giving the expected leading dependence on $a$ for Wilson fermions close to the chiral limit:
\beq\label{eq:SUSY_limit_fit_function}
\left(\wzero \Mgg\right)(\infty,a, m_\pi) = (\wzero \Mgg)(\infty) + c_1 \frac{a}{\sqrt{8\tone}} + c_2 (8\tone)m^2_\pi.
\eeq
Note that, as derived in Ref.~\cite{Munster:2014cja} in Partially Quenched Chiral Perturbation Theory, loop effects are expected to exactly cancel out in this case, so that no chiral log appears in the chiral expansion. Moreover, higher-order terms in the adjoint-pion mass appear to be negligible in the explored range of parameters. The quantity $(\wzero \Mgg)(\infty)$ represents the $\SU(\infty)$ gluino-glue mass in the SUSY limit, whose value, plotted as a starred point in Fig.~\ref{fig:SUSY_limit}, is given by:
\beq\label{eq:final_largeN_res}
(\wzero \Mgg)(\infty) = 1.21(11), \qquad \qquad \text{($N=\infty$, SUSY limit)}.
\eeq
Using Eqs.~\eqref{eq:ratio_w0_t1_largeN}-\eqref{eq:ratio_t0_t1_largeN}, we are able to also express our final result in units of the popular gradient flow scale $\sqrt{8\tzero}$:
\beq
(\sqrt{8\tzero}\Mgg)(\infty) &=& 3.36(33), \qquad \qquad \text{($N=\infty$, SUSY limit)}.
\eeq

It should be noted that in Fig.~\ref{fig:SUSY_limit} we reported both the gluino-glue mass determinations and the best fit curve in the continuum limit, namely, subtracting the finite-lattice-spacing $c_1 \frac{a}{\sqrt{8\tone}}$ correction. Finally, we also verified that adding a further mixed term to the fit function of the form $c_3 \left(\frac{a}{\sqrt{8\tone}}\right) \times \left[(8\tone)m^2_\pi\right]$ results in a value of $c_3$ compatible with zero within uncertainties, $c_3=0.21(20)$, and yields a final result for $\wzero \Mgg$ in the SUSY limit which is compatible with~\eqref{eq:final_largeN_res} within a larger uncertainty, $\wzero \Mgg = 1.61(40)$.

To conclude, we will express our final result for $\Mgg$ in more conventional units of the continuum theory, namely, the $\Lambda$-parameter in the Novikov--Shifman--Vainshtein--Zakharov (NSVZ) scheme~\cite{Novikov:1983ee,Shifman:1986zi}:
\beq\label{eq:SUSY_Lambda_param}
\Lambda_{\NSVZ}^3 = \frac{\mu^3}{b_0\lambda_{\rm t}(\mu)}\exp\left(-\frac{8\pi^2}{\lambda_{\rm t}(\mu)}\right), \qquad \qquad b_0 = 3/(4\pi)^2,
\eeq
where $\lambda_{\rm t}(\mu) = N g^2(\mu)$ is the renormalized running 't Hooft coupling in the NSVZ scheme. Note that the definition in Eq.~\eqref{eq:SUSY_Lambda_param} follows the usual QCD convention, and differs from the one typically employed in the SUSY literature by the extra factor of $1/b_0$. Using our non-perturbative $N=\infty$ result obtained in~\cite{Bonanno:2024bqg},
\beq
\sqrt{8\tone} \Lambda_{\NSVZ} = 0.231(15), \qquad \qquad (N=\infty).
\eeq
and $\wzero/\sqrt{8\tone} = 0.586(10)$, we can eliminate the reference scale $\wzero$ and express our result for the large-$N$ gluino-glue mass in terms of the SUSY scale $\Lambda_{\NSVZ}$:
\beq
\frac{\Mgg}{\Lambda_\NSVZ} &=& 8.94 \pm 1.01, \qquad \qquad (N=\infty).
\eeq
We recall that, as discussed in Refs.~\cite{Finnell:1995dr,Armoni:2003yv}, the $\Lambda$-parameter in the NSVZ scheme is equal to the one in the Dimensional Reduction ($\overline{\mathrm{DR}}$) scheme: $\Lambda_{\NSVZ}=\Lambda_{\scriptscriptstyle{\overline{\mathrm{DR}}}}$.

\subsection{Comparison with previous results at finite  \texorpdfstring{$N$}{N}}\label{sec:Ndep}

This section is devoted to the comparison of our large-$N$ determination with previous lattice non-perturbative results of $\wzero \Mgg$ in the SUSY limit obtained for finite values of $N$.

The $\SU(3)$ value in the SUSY limit:
\beq
\wzero \Mgg = 1.042(46), \qquad \qquad (N=3),
\eeq
was obtained from a chiral-continuum extrapolation of the data reported in the Supplemental Material of Ref.~\cite{Ali:2019agk} assuming leading $\mathcal{O}(a^2)$, $\mathcal{O}(m_\pi^2)$ and $\mathcal{O}(m_\pi^2a^2)$ corrections. This number matches with the extrapolation shown in Fig.~1 of Ref.~\cite{Ali:2019agk}.\footnote{We note that there is a mismatch between the values of $\wzero \Mgg$ in the SUSY limit reported in Tab.~1 and displayed in Fig.~1 of Ref.~\cite{Ali:2019agk}.} It is important to stress that, in Ref.~\cite{Ali:2019agk}, the scale $\wzero$ is defined consistently with our definition in Eq.~\eqref{eq:ref_scale_w0}.

To obtain the value for $N=2$ instead we must combine the result of Ref.~\cite{Ali:2019gzj} for $\wzero(c=0.15) \Mgg = 0.93(6)$ with the result for $\wzero(c=0.1)/\wzero(c=0.15) = 0.8815(19)$ of Ref.~\cite{Bergner:2014ska} in order to express it in terms of the same scale defined in Eq.~\eqref{eq:ref_scale_w0}. In the end one finds:
\beq
\wzero \Mgg = 0.823(56), \qquad \qquad (N=2).
\eeq

We summarized all available non-perturbative results for the gluino-glue mass in the SUSY limit in Tab.~\ref{tab:final_res}. We have also included the determinations for $N=2$ and 3 in Fig.~\ref{fig:SUSY_limit} for comparison with our large-$N$ data. As it can be observed, our $N=\infty$ result turns out to be close but larger compared to $N=2,3$ determinations, thus confirming the trend already observed for finite-$N$ results, which already hinted towards a growth of $\Mgg$ as a function of $N$.

\begin{table}[!t]
\small
\begin{center}
\begin{tabular}{|c|c|}
\hline
$N$ & $\wzero \Mgg$ \\
\hline
2 & 0.823(56)\\
3 & 1.042(46)\\
$\infty$ & 1.21(11)\\
\hline
\end{tabular}
\end{center}
\caption{Collection of the non-perturbative determination of the gluino-glue mass $\Mgg$ in units of the reference hadronic scale $\wzero$. The $N=2$ determination comes from Refs.~\cite{Bergner:2014ska,Ali:2019gzj}, the $N=3$ from Ref.~\cite{Ali:2019agk}.}
\label{tab:final_res}
\end{table}

\begin{figure}[!t]
\centering
\includegraphics[scale=0.62]{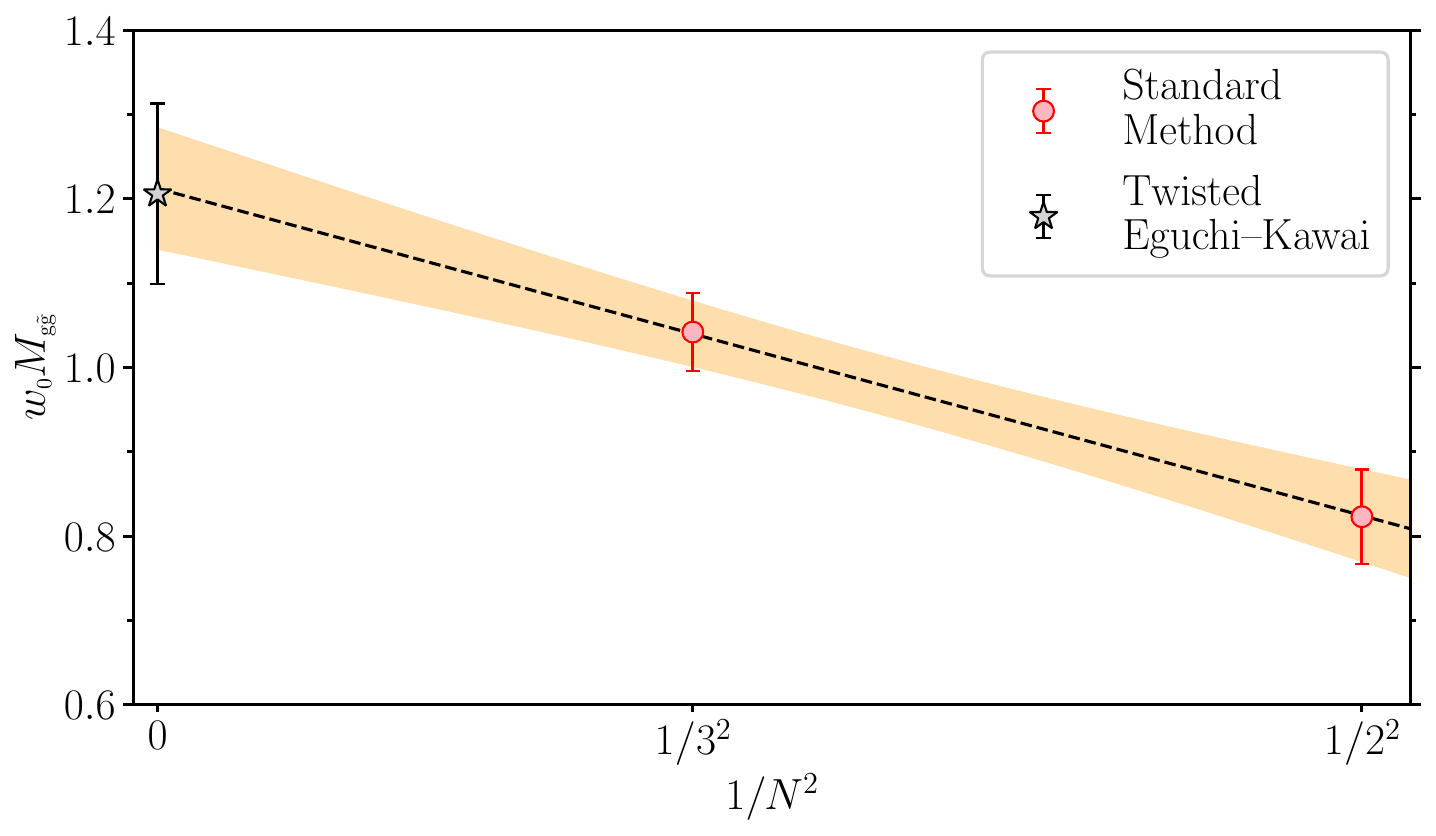}
\caption{Best fit of the results for $\wzero \Mgg$ as a function of $N$ reported in Tab.~\ref{tab:final_res} according to a fit function $f(N)=A+B/N^2$. Best fit results are reported in Eq.~\eqref{eq:final_res_Ndep}.}
\label{fig:Mgg_Ndep}
\end{figure}

Standard large-$N$ arguments in the presence of adjoint fermions predict that, for $N$ sufficiently large, any physical observable can be expanded around $1/N=0$ in a power series of $1/N^2$. Several recent lattice results obtained in non-supersymmetric large-$N$ SU($N$) Yang--Mills theory pointed out that the $N$-dependence of several physical quantities turns out to be pretty smooth, and that large-$N$ scaling arguments already work for relatively small values of $N$, see, e.g., Refs.~\cite{Lucini:2001ej,Lucini:2004my,Lucini:2010nv,Lucini:2012gg,Bali:2013kia,Bonati:2016tvi,Ce:2016awn,DeGrand:2016pur,Hernandez:2019qed,Bennett:2020hqd,DeGrand:2020utq,Hernandez:2020tbc,Bonanno:2020hht,DeGrand:2021zjw,Athenodorou:2021qvs,Bonanno:2022yjr,Bennett:2022gdz,Bonanno:2023ypf,DeGrand:2023hzz,Bonanno:2023ple,Bonanno:2023hhp,Bonanno:2024ggk}. However, it would be too optimistic to expect large-$N$ arguments to work down to $N=2$, where sizable corrections to the leading large-$N$ behavior are in principle to be expected.

With all these caveats in mind, once our large-$N$ value is plotted with available finite-$N$ determinations as a function of $1/N^2$, we observe that the resulting $N$-dependence turns out to be rather mild, see Fig.~\ref{fig:Mgg_Ndep}. In particular, it is possible to perform a naive fit just including the leading $1/N^2$ correction to the $N=\infty$ value, resulting in a very small chi-squared and in the result:
\beq\label{eq:final_res_Ndep}
(\wzero \Mgg)(N) = 1.212(71) - \frac{1.55(41)}{N^2},
\eeq
where the first coefficient is in very good agreement with our direct large-$N$ determination~\eqref{eq:final_largeN_res}. Although this result should be taken with a grain of salt, it nonetheless can be regarded as an indication that higher-order correction in $1/N^2$ are not excessively large, and cannot be observed within current statistical uncertainties.

\section{Conclusions}\label{sec:conclu}

In this work we have performed the first non-perturbative computation of the mass of the gluino-gluon bound state in large-$N$ $\mathcal{N}=1$ SUSY Yang--Mills theory. This state is a member of the triply-degenerate supermultiplet expected to constitute the ground state of the theory, based on supersymmetry arguments.

After extrapolation towards the thermodynamic and SUSY-restoring limits, we found, in terms of the SUSY $\Lambda$-parameter in the NSVZ scheme:
\beq
\frac{\Mgg}{\Lambda_\NSVZ} = 8.94 \pm 1.01 \qquad \qquad (N=\infty).
\eeq
For completeness, we also report our final $\SU(\infty)$ result in terms of the reference hadronic scales $\wzero$ and $\sqrt{8\tzero}$ typically used in lattice computations:
\beq
\wzero\Mgg &=& 1.21(11), \quad \quad (N=\infty),\\
\sqrt{8\tzero}\Mgg &=& 3.36(33), \quad \quad (N=\infty).
\eeq

Our $\SU(\infty)$ result turns out to be remarkably close but larger than previous non-perturbative determinations of $\wzero \Mgg$ obtained at finite values of $N$, and confirms the trend already observed for $N=2$ and 3 that $\Mgg$ grows as $N$ is increased. Once all available data for $\Mgg$ are compared as a function of $1/N^2$, the resulting $N$-dependence turns out to be pretty smooth, hinting towards a scenario where higher-order $1/N^2$ corrections are not excessively large, and cannot be observed within current statistical errors.

Assuming supersymmetry, our result for $(\Mgg/\Lambda_\NSVZ)(N=\infty)$ also expresses the mass of the lightest supermultiplet of large-$N$ SUSY Yang--Mills, as the gluino-gluon bound state is expected to be degenerate with the lightest scalar and pseudo-scalar bound states, thus constituting a prediction for the mass gap of the large-$N$ SUSY theory. In the next future it would be interesting to confirm this by explicitly calculating the ground-state masses of the candidate superpartners at large-$N$ using the TEK reduced model. To this end, by virtue of large-$N$ volume independence, one could adopt a spatially-reduced lattice with a temporal extension $N_t > 1$, which would allow to compute double-trace time-correlators and use standard lattice spectroscopy methods to address the mass extraction of the other members of the supermultiplet.

\section*{Acknowledgements}
We thank P.~Butti for collaboration in the early stages of this study. This work is partially supported by the Spanish Research Agency (Agencia Estatal de Investigaci\'on) through the grant IFT Centro de Excelencia Severo Ochoa CEX2020-001007-S and, partially, by the grant PID2021-127526NB-I00, both of which are funded by MCIN/AEI/10.13039/ 501100011033. K.-I.~I.~is supported in part by MEXT as ``Feasibility studies for the next-generation computing infrastructure''. M.~O.~is supported by JSPS KAKENHI Grant Number 21K03576. Numerical calculations have been performed on the \texttt{Finisterrae~III} cluster at CESGA (Centro de Supercomputaci\'on de Galicia). We have also used computational resource of Cygnus at Center for Computational Sciences, University of Tsukuba.

\providecommand{\href}[2]{#2}\begingroup\raggedright\endgroup

\end{document}